\documentclass[conference,usenatbib]{basi}
\usepackage[T1]{fontenc}
\usepackage[british]{babel}
\usepackage[varg]{txfonts}
\usepackage{rotating}
\usepackage{dcolumn}
\begin{document}
\title[Synchrotron from recurrent novae]{Modelling the synchrotron light curves in
recurrent novae V745 Scorpii and RS Ophiuchi}
\author[N.~G.~Kantharia et~al.]%
       {N.~G.~Kantharia$^1$\thanks{email: \texttt{ngk@ncra.tifr.res.in}},
       P.~Dutta$^{2}$, N.~Roy$^3$,
       G.~C.~Anupama$^{4}$, A.~Chitale,
      \newauthor C.~Ishwara-Chandra$^{1}$, T.~P.~Prabhu$^4$,
       N.~M.~Ashok$^{5}$, D.~P.~K.~Banerjee $^5$\\
       $^1$NCRA-TIFR, Pune,~$^2$IISER, Bhopal,~
       $^3$Dept. of Physics, IIT, Kharagpur~$^4$IIA, Bangalore~$^5$PRL, Ahmedabad \\
}

\pubyear{2015}
\volume{12}
\pagerange{\pageref{firstpage}--\pageref{lastpage}}

\date{Received --- ; accepted ---}

\maketitle
\label{firstpage}

\begin{abstract}
In this paper, we present the synchrotron light curve at 610 MHz from the recurrent
nova V745 Sco following its outburst on 6 February 2014.  The 
system has been detected and periodically monitored with the 
Giant Metrewave Radio Telescope 
(GMRT) since 9 February 2014 as part of the Galactic Nova with GMRT (GNovaG)
project. The light curves are well fit by a model of
synchrotron emitting region obscured by foreground thermal gas which eventually becomes
optically thin to the low GMRT frequencies.  We present the model fit
to the 2014 data on V745 Sco and discuss it alongwith the model fit to the 1.4 GHz data of
the recurrent nova RS Ophiuchi following its outburst in 1985.
\end{abstract}


\section{Introduction}\label{s:intro}

Novae are cataclysmic variables comprising a white dwarf primary star and
a main sequence or evolved secondary star.   The primary accretes matter from the
secondary leading to thermonuclear outbursts on the surface of the white dwarf.
Novae brighten by several optical magnitudes
and the ones which periodically undergo outbursts are referred to as recurrent novae.  
Emission from the system due to different physical processes is
detected in bands ranging from $\gamma$-rays to radio waves.  The emission
at the long radio wavelengths ($ > 21$cm) is dominantly due to the synchrotron process.
The energy released in a nova outburst is $10^{38-43}$ ergs and the
synchrotron radio luminosity is about $10^{20}$ erg~s$^{-1}$~Hz$^{-1}$ 
\citep{2012BASI...40..311K}.

\section{Outburst in the recurrent nova V745 Sco in 2014}

\begin{figure}
\centerline{\includegraphics[width=4.6cm,angle=-90]{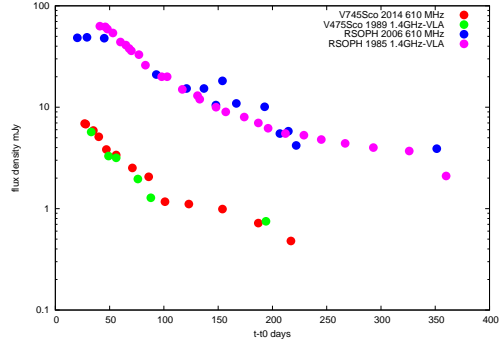}}
\caption{Synchrotron emission light curves of recurrent novae: V745 Sco
(outbursts 1989, 2014: Kantharia et al. 2015 (submitted to MNRAS)) and RS Ophiuchi (outbursts 1985: \citet{1986ApJ...305L..71H}, 
2006: \citet{2007ApJ...667L.171K}).   The 610 MHz data is obtained using GMRT. }

\label{fig1}
\end{figure}
Following its outburst in 2014, 
V745Sco was detected at 610 MHz using GMRT \citep{2014ATel.5962....1K}.
The light curves of V745 Sco following its outburst in 2014 and 1989 and 
of RS Ophiuchi following its outbursts in 1985 and 2006 are shown in Fig. \ref{fig1}.  
Notice the similar evolution of the radio light curves from multiple outbursts. 
A model consisting of a synchrotron emitting
shell which is initially obscured by foreground thermal gas leading to delayed turnon
of low radio frequency synchrotron emission well fits the observed light curve.   We use
the model parametrized by \citet{2002ARA&A..40..387W} to explain supernova light curves and
have included only foreground absorption due to clumpy and uniform components.
The best model gives a turn-on on day 9.5 and peak emission on 
day 23 for the 610 MHz data on V745 Sco in 2014.  
A similar model is found to fit the VLA 1.4 GHz light curve obtained
following an outburst in the recurrent nova RS Ophiuchi in 1985.  More details
are presented in a research paper submitted to MNRAS. 

\section*{Acknowledgements}
We thank the staff of the GMRT that made these observations possible. GMRT is run by NCRA,
a centre of TIFR.

\end{document}